\documentclass[prl,
twocolumn,
aps,prc,showpacs,
tightenlines,nofootinbib]{revtex4}
\usepackage{epsf}
\usepackage{amsmath}
\usepackage{nicefrac}
\usepackage{esint}
\def\beq{\begin{equation}}
\def\eeq{\end{equation}}
\def\bea{\begin{eqnarray}}
\def\eea{\end{eqnarray}}
\def\beqa{\begin{equation}\begin{array}{l}}
\def\eeqa{\end{array}\end{equation}}
\def\eqlab#1{\label{eq:#1}}
\def\figlab#1{\label{fig:#1}}

\def\eref#1{(\ref{eq:#1})}
\def\Eqref#1{Eq.~(\ref{eq:#1})}

\def\Figref#1{Fig.~\ref{fig:#1}}

\def\barr{\left(\begin{array}{c}}
\def\earr{\end{array}\right)}
\def\bmat{\left(\begin{array}{cc}}
\def\emat{\end{array}\right)}
\def\al{\alpha}
\def\be{\beta}
\def\ga{\gamma} 
 \def\De{\Delta}
\def\veps{\varepsilon}  

\def\la{\lambda}

\def\si{\sigma} 
\def\th{\theta}  
\def\w{\omega}

\def\dd{{\rm d}}
\def\pa{\partial}

\def\pa{\partial}

\def\nn{\nonumber}

\def\mathscr{\mathcal}

\def\re{\mbox{Re}}
\def\im{\mbox{Im}}
\def\3d{3-D}

\begin{document}
\preprint{}

\title{Sum rules for light-by-light scattering}

\author{Vladimir Pascalutsa}
%\email{vladipas@kph.uni-mainz.de}
\affiliation{Institut f\"ur Kernphysik, Johannes Gutenberg Universit\"at, Mainz D-55099, Germany}

%\affiliation{European Centre for Theoretical Studies in Nuclear Physics and Related Areas (ECT*), \\
%Villa Tambosi, Villazzano (Trento),
%I-38050 TN, Italy}

\author{Marc Vanderhaeghen}
%\email{marcvdh@kph.uni-mainz.de}
\affiliation{Institut f\"ur Kernphysik, Johannes Gutenberg Universit\"at, Mainz D-55099, Germany}

\date{\today}

\begin{abstract}
We derive a set of sum rules for the light-by-light scattering and fusion: 
$\gamma\gamma \to all$, and verify them in lowest order QED calculations. 
A prominent implication of these sum rules is the superconvergence of
the helicity-difference total cross-section for photon fusion, which in the 
hadron sector reveals an intricate cancellation between the pseudoscalar and
tensor mesons. 
An experimental verification of superconvergence of the polarized photon fusion
into hadrons is called for, but will only be possible at $e^+ e^-$ and $\ga\ga$ colliders 
with  both beams polarized. We also show how the sum rules can be used
to measure various contributions to the low-energy light-by-light scattering.
\end{abstract}


\maketitle
\thispagestyle{empty}

%\section{Introduction}

Light-by-light scattering is a well-known phenomenon  of purely quantum origin  
\cite{Heisenberg:1935qt,Karplus:1950zza}, which has been observed (albeit indirectly)
at nearly all high-energy colliders, see \cite{Budnev:1974de,Brodsky:2005wk,Klempt:2007cp}
for reviews. The inclusive unpolarized cross-section for 
the photon fusion  ($\gamma\gamma \to  all$)
is fairly well measured for energies between 1 and 200 GeV \cite{Amsler:2008zzb},
but not much is known about the polarized cross-sections, 
 \begin{subequations}
  \bea
 \De \si(\w^2) & \equiv & \si_2(\w^2) - \si_0(\w^2), \\
  \tau(\w^2) & \equiv & \si_{||}(\w^2) - \si_\perp(\w^2),
 \eea 
  \end{subequations}
where the subscripts indicate the photon polarizations ($0$ or $2$ show the total helicity of the
circularly polarized photons,  while $||$ or $\perp$ show if the linear photon polarizations
are parallel or perpendicular);  
$\w$ is the photon energy in collider kinematics. To access these observables at 
high-luminosity colliders, the beams need to be polarized  which has not been realized so far. 

In this Letter we derive sum rules which relate integrals of the total polarized and unpolarized
cross sections to the low-energy structure of light-by-light scattering.
The sum rule for the helicity-difference cross section is particularly simple:
 \beq
 \eqlab{theSR}
 \int_{0}^\infty \!\! \dd \w \, \frac{\De \si(\w^2)}{\w} = 0\,.
 \eeq
This sum rule can be inferred \cite{Roy:1974fz,Brodsky:1995fj} from the
 Gerasimov-Drell-Hearn (GDH) sum rule~\cite{GDH} applied to the $\ga\ga$ process. 
 The other two sum rules to be presented here determine the low-energy
 constants of the Euler-Heisenberg Lagrangian in terms of the integrals
 over the linearly-polarized cross-section of photon fusion.  Contrary 
 to the sum rule \eref{theSR}, those two sum rules cannot entirely be inferred from
 the Compton-scattering sum rules.
 
 Nevertheless, the general strategy of the derivation is, of course, the same.
As in the GDH case, we only need
the general assumptions about analyticity, crossing-symmetry, unitarity, 
and the low-energy limit of the forward elastic-scattering amplitude. The only difference  
is that instead of scattering the photon on a massive target (Compton scattering), we consider
the elastic light-by-light scattering ($\gamma \gamma \to \gamma \gamma$).
Denoting the corresponding Feynman and helicity amplitudes as, respectively, $\mathcal{M}$ and $M$, their relation is
\bea
\eqlab{genexp}
M_{\la_1\la_2\la_3\la_4} &=& \veps^{\ast\mu_4}_{\la_4}(\vec{q}_4) \, \veps^{\ast\mu_3}_{\la_3} (\vec{q}_3)\, 
\veps^{\mu_2}_{\la_2}(\vec{q}_2)\,  \veps^{\mu_1}_{\la_1}(\vec{q}_1) \nn\\
& & \times \,  \mathcal{M}_{\mu_1\mu_2\mu_3\mu_4} ,
\eea
where $\veps(\vec{q})$ are the photon polarization 4-vectors, $\la$'s are the helicities; for 
real photons traveling along the $z$ axis, i.e.\ $\vec{q} = (0,0,\w)$, the polarization vectors are
$\veps_{\la}(\pm \vec{q}) = 2^{-\nicefrac12} ( 0, \mp \la, -i , 0)$.
The Mandelstam variables are defined as
$s=(q_1+q_2)^2 = 4 \w^2$, $t=(q_1-q_3)^2$, $u=(q_1-q_4)^2$, with $q_i$ the
photon 4-momenta. 

In the forward kinematics, where $q_3=q_1$, $q_4=q_2$,
and hence $t=0$, $u=-s$, the general Lorentz structure of the Feynman amplitude is given by:
\bea
\mathcal{M}_{\mu_1\mu_2\mu_3\mu_4} & =&  A(s) \, g_{\mu_4\mu_2} g_{\mu_3\mu_1} 
+B(s)\, g_{\mu_4\mu_1} g_{\mu_3\mu_2} \nn\\
&  +&  C(s) \, g_{\mu_4\mu_3} g_{\mu_2\mu_1} \,,
\eea
where $g_{\mu\nu}$ is the Minkowski metric. Crossing symmetry (under $1 \leftrightarrow 3$, or 
$2 \leftrightarrow 4$) means in this case
\bea
\mathcal{M}_{\mu_1\mu_2\mu_3\mu_4} & =&  A(u) \, g_{\mu_4\mu_2} g_{\mu_3\mu_1} 
+B(u)\, g_{\mu_4\mu_3} g_{\mu_1\mu_2} \nn\\
&  +&  C(u) \, g_{\mu_1\mu_4} g_{\mu_3\mu_2} \,,
\eea 
hence $A(-s) = A(s)$, $B(-s) = C(s)$.
 As a result there are three independent nonvanishing helicity amplitudes:
 \bea
  M_{++++}(s) &=& A(s) + C(s),\nn\\
  M_{+-+-}(s) &=& A(s) + B(s),\\
  M_{++--}(s) &=& B(s) + C(s),\nn
 \eea
 satisfying the  following crossing relations:
$ M_{+-+-}(s) =   M_{++++}(-s)$, and
$ M_{++--}(s) = M_{++--}(-s)$.

We next turn to the causality constraint, which implies that the above functions
are analytic functions of $s$ everywhere in the complex $s$ plane  except along the real axis.
For the amplitudes,
 \begin{subequations}
 \bea
f^{(\pm)}(s) &=&  M_{++++}(s)  \pm M_{+-+-}(s) ,\\
g(s) &=&  M_{++--}(s),
\eea
 \end{subequations}
the analyticity  infers dispersion relations:
\beq
\re\, \left\{\begin{array}{l} f^{(\pm)} (s)\\
g (s)\end{array} \right\}  = \frac{1}{\pi} \fint\limits_{-\infty}^{\infty} \! \frac{\dd s' }{s'-s} \, 
 \im\, \left\{\begin{array}{l} f^{(\pm)} (s')\\
g (s')\end{array} \right\}\,,
\eeq
where $ \fint$ indicates the principal-value integration. These relations hold as long as
the integral converges, and otherwise subtractions are needed.
Because $f^{(\pm)}(-s)=\pm \,f^{(\pm)}(s)$ and $g(-s)=g(s)$, 
we can express the right-hand side as an integral over positive $s$
only:
 \begin{subequations}
 \bea
\re\, \left\{\begin{array}{l} f^{(+)} (s)\\
g (s)\end{array} \right\} & = & \frac{2}{\pi}  \fint\limits_{0}^{\infty} \frac{\dd s' \, s'}{s^{\prime 2}-s^2}\, 
 \im\, \left\{\begin{array}{l} f^{(+)} (s')\\
g (s')\end{array} \right\}\,,\\
\re\, f^{(-)} (s) & = & -\frac{2s}{\pi} \fint\limits_{0}^{\infty} \dd s' \, \frac{  \im\, f^{(-)} (s')}{s^{\prime 2}-s^2}\,.
\eea
\end{subequations}
In the physical region ($s\geq 0$),  the optical theorem relates 
the imaginary part of these amplitudes to the total absorption cross-sections with
definite polarization of the initial $\gamma \gamma$ state:
 \begin{subequations}
\bea
\im\, f^{(\pm)}(s) &=&  - \frac{s}{8} \,[ \,\si_0(s)  \pm \si_2(s) \,] ,\\
\im\, g(s) &=&  - \frac{s}{8} \,[ \,\si_{||}(s)  - \si_\perp (s) \,] .
\eea
\end{subequations}
Substituting these expressions in the above dispersion relations we obtain:
 \begin{subequations}
 \eqlab{srules}
 \bea
\re\, f^{(+)} (s) & = & -\frac{1}{2\pi}\fint\limits_{0}^{\infty} \dd s'\, s^{\prime 2}\,\frac{\si(s')}{s^{\prime 2}-s^2}\,,
\eqlab{srulesa}\\
\re\, f^{(-)} (s) & = & -\frac{s}{4\pi} \fint\limits_{0}^{\infty} \dd s' \,\frac{ s'  \, \De\si (s')}{s^{\prime 2}-s^2}\,,\eqlab{srulesb}\\
\re\, g (s) & = & -\frac{1}{4\pi} \fint\limits_{0}^{\infty} \dd s' \,s^{\prime 2}  \,\frac{ 
\si_{||}(s')  - \si_\perp (s')}{s^{\prime 2}-s^2}\,,
\eqlab{srulesc}
\eea
\end{subequations}
where $\si=(\si_0+\si_2)/2=(\si_{||} + \si_\perp)/2$ is the unpolarized 
total cross section. 

We next recall that gauge invariance and discrete symmetries constrain the low-energy
photon-photon interaction to the Euler-Heisenberg form \cite{Heisenberg:1935qt}, given by the following Lagrangian density:
\beq
\eqlab{EHlagr}
\mathcal{L}_{\mathrm{EH}} = c_1 (F_{\mu\nu}F^{\mu\nu})^2 + c_2 (F_{\mu\nu}\tilde F^{\mu\nu})^2,
\eeq
where $F_{\mu\nu} = \pa_\mu A_\nu - \pa_\nu A_\mu$, $\tilde F^{\mu\nu} = 
\veps^{\mu\nu\al\be} \pa_\al A_\be$. Evidently, the low-energy expansion of the above amplitudes begins with $\w^4$. Expanding the left-hand side and right-hand side of \Eqref{srules} in powers of $s$
and matching them at each order yields a number of sum rules. At 0th order in $s$
we would find
 \bea
0 & = &  \int\limits_{0}^{\infty} \dd s \, \Big[ \si_{||} (s) \pm \si_\perp(s) \Big]\,,
\eea
which cannot work for ``+"  since the unpolarized cross-section is a positive-definite quantity.
Empirically $\si$ shows a slowly rising behavior at large $s$ and the integral diverges. 
The assumption of an unsubtracted dispersion relation is violated in this case. 
For the ``$-$" case the sum rule is broken too, cf. ~\cite{Gerasimov:1973ja} and references
therein.

 At the  first and second orders we find, respectively:
\begin{subequations}
 \eqlab{s0rules}
 \bea
0 & = &  \int\limits_{0}^{\infty} \dd s\,  \frac{ \De\si (s)}{s}\,,
\eqlab{s0rulesb} \\
c_1\pm c_2 & = & \frac{1}{8\pi }\int\limits_{0}^{\infty} \dd s\,  \frac{ \si_{||} (s) \pm \si_\perp(s)}{s^2}\,.
 \eqlab{s0rulesc}
\eea
\end{subequations}

\begin{figure}[t]
\centerline{\epsfclipon  \epsfxsize=8cm%
  \epsffile{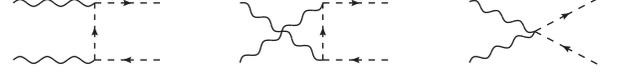} 
}
\caption{Pair production in scalar QED.
% (Drawn using JaxoDraw.)
}
\figlab{scalar}
\end{figure}

%\begin{widetext}
The first of these sum rules is just the superconvergence relation \eref{theSR}. 
Although $\De \si$ is not known experimentally,  we can test it in (field) theory.
In scalar QED, for example, to leading order in the fine-structure constant $\al$,
the production of a scalar-antiscalar pair ($\ga\ga\to S^+ S^-$) is described by the
Feynman graphs in \Figref{scalar}, which give rise to the following helicity-difference cross-section:
\bea
 && \De \si^{(S^+ S^-)} (s) = \frac{4\pi \al^2}{s} \, \th(s-4m^2) 
 \label{eq:scalarsr} \\
  && \quad \times \, \left( \sqrt{1-\frac{4m^2}{s}} 
 -  \frac{8m^2}{s}\, \mathrm{arctanh} \sqrt{1-\frac{4m^2}{s} }
  \, \right),\nn
 \eea
 where $m$ is the mass of the scalar. It is easy to see that this expression verifies
 \Eqref{s0rulesb}.
A similar check in spinor electrodynamics works out as well. We shall demonstrate it
for the case of virtual photons.
%\medskip
%\end{widetext}

The arguments used above in deriving the sum rules of \Eqref{srules} will equally hold for 
space-like virtual photons ($q_1^2  < 0 $, $q_2^2 < 0 $), if  written
in a variable which reflects under crossing, e.g. $\nu = s-q_1^2-q_2^2$. 
 %As the result we obtain the following generalization:
%\beq
%  \fint\limits_{0}^{\infty} \dd s\,  \frac{ \De\si (s,q_1^2,q_2^2)}{s-q_1^2-q_2^2}=0\,,
%\eeq
% where $\De\si (s,q_1^2,q_2^2)$ is helicity-difference total cross-section of
% $\ga^\ast \ga^\ast \to all$. 

When one of the photons virtuality is large, entering the deep inelastic scattering (DIS)
regime, this sum rule  converts to a 
sum rule for the photon structure function $g_1^\gamma$, 
cf.~\cite{Bass:1998bw}.

 \begin{figure}[t]
\centerline{\epsfclipon  \epsfxsize=5.5cm%
  \epsffile{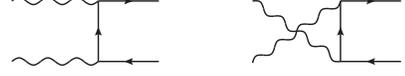} 
}
\caption{$e^+ e^-$ production in QED.
% (Drawn using JaxoDraw.)
}
\figlab{spinor}
\end{figure}

Let us now consider how the superconvergent sum rule  \eref{s0rulesb} works in QED in
the case of one space-like virtual photon ($q^2 < 0$).
 At leading order in $\al$ only the 
 tree-level  fermion-pair production ($\ga^\ast \ga \to f \bar f$) contributes
 to $\De \si$, see 
 \Figref{spinor}.
The expressions for the corresponding helicity difference cross-section exists
in the literature\footnote{We only reinstall a missing overall factor of 2.
Note also that in the notation of \cite{Budnev:1974de}, 
$\De\si  = -\tau_{TT}^a$.}~\cite{Budnev:1974de}:
%\begin{widetext}
\bea
 && \De \si^{(\ga^\ast\ga\to f \bar f)} (s,q^2,0) =  \frac{8\pi \al^2}{(s-q^2)^2} \, \th(s-4m^2)\\
 &&\, \times  \left\{ 
  -(3 s+q^2)\sqrt{1-\frac{4m^2}{s}}  + 2 (s+q^2)\, \mathrm{arccos} \frac{s^{\nicefrac12}}{2}
  \, \right\}. \nn
 \eea
  In \Figref{xsect} we plot this cross section as a function of energy, for three different values
  of $q^2$. One sees that in all cases the low- and high-energy contributions cancel. 
% \end{widetext}
The result
 \beq
  \int\limits_{4 m^2}^{\infty} \dd s\,  \frac{ \De\si^{(\ga^\ast\ga\to f \bar f)} (s,q^2,0)}{s-q^2}=0\,
\eeq
is easily verified for any $q^2 < 4 m^2$.

 \begin{figure}[b]
\centerline{\epsfclipon  \epsfxsize=6.7cm%
  \epsffile{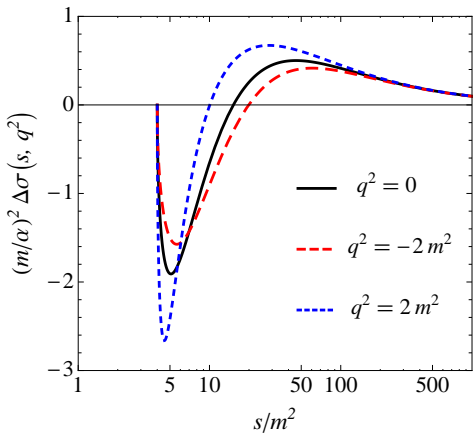} 
}
\caption{Helicity-difference cross section of  $\ga^\ast\ga\to f \bar f$ in  QED
 at leading order,
for different photon virtualities.
% (Drawn using JaxoDraw.)
}
\figlab{xsect}
\end{figure}

While the role of this sum rule in QED becomes fairly clear in these perturbative calculations,
in quantum chromodynamics (QCD), in its non-perturbative regime, it is far less obvious.  
We can gain some
insight by looking at  individual contributions to the $\gamma \gamma\to hadrons$ cross-sections.

The high-energy behavior of $\Delta \sigma$ is determined by a $t$-channel 
exchange of unnatural parity and is expected from Regge theory -- in the absence of fixed pole singularities -- to drop as $1/s$ or faster~\cite{Budnev:1971sz}. We therefore expect the 
sum rule \eref{s0rulesb} to converge. 
The dominant features of the $\gamma \gamma$ to multihadron production  comes firstly from the 
Born terms in the $\pi^+ \pi^-$ (or $K^+ K^-$) channels. These contributions are given by 
Eq.~(\ref{eq:scalarsr}) and hence each separately obeys the sum rule.
The largest contributions in the hadronic sector are thus expected to come from the resonance production:
$\gamma \gamma \to M$, with $M$ being a meson. It is highly nontrivial to see how the
sum rule is saturated in this case. 

As two real photons do not 
couple to a $J^P$ = $1^-$ or $1^+$ state due to the Landau-Yang theorem, one expects the 
dominant contribution to come from scalar, pseudoscalar, and tensor mesons. 
One can express the $\gamma \gamma \to M$ cross section for a meson with spin $J$, mass $m_M$, and total width $\Gamma_{tot}$, using a Breit-Wigner parametrization, in terms of the 
decay width $\Gamma_{\gamma \gamma}^{(\Lambda)}$ of the meson 
into two photons of total helicity $\Lambda = 0, 2$, as   
\begin{eqnarray}
\sigma^{\gamma \gamma \to M}_\Lambda (s) &=& (2 J + 1) \, 16 \pi 
 \frac{\Gamma_{\gamma \gamma}^{ (\Lambda)} \, \Gamma_{tot}}{(s - m_M^2)^2 + \Gamma_{tot}^2 m_M^2} \nonumber \\
&\approx& (2 J + 1) 16 \pi^2 
\, \frac{\Gamma_{\gamma \gamma }^{(\Lambda)} }{m_M} \, \delta(s - m_M^2),
\label{eq:mesgaga}
\end{eqnarray}
where the last line is obtained in the narrow resonance approximation. 
For the pseudoscalar mesons, which can only contribute to the helicity-zero cross section, the narrow resonance approximation is very accurate and allows to quantify their contribution as shown 
in Table~\ref{table_ps}. For the pion, this value is entirely driven by the chiral anomaly, which allows 
the expression of the $\pi^0$ contribution to the sum rule as $- \alpha^2 / (4 \pi f_\pi^2)$, with $f_\pi = 92.4$~MeV the pion decay constant.

To compensate the large negative contribution to the sum rule from pseudoscalar mesons, 
one needs to have an equivalent strength  in the helicity-two cross section, $\sigma_2$.
The dominant feature of the helicity-two cross section in the resonance region  
arises from the multiplet of tensor mesons $f_2(1270)$, $a_2(1320)$, and $f_2^\prime(1525)$. 
Measurements at various $e^+ e^-$ colliders, notably recent high statistics measurements 
by the BELLE Collaboration of the $\gamma \gamma$ cross sections to 
$\pi^+ \pi^-$~\cite{Mori:2007bu},  $\pi^0 \pi^0$~\cite{Uehara:2008pf}, $\eta \pi^0$~\cite{etapi:2009cf}, 
and $K^+ K^-$~\cite{Abe:2003vn} channels have allowed accurate confirmation of their parameters. 
As these tensor mesons were also found to be relatively well described by Breit-Wigner resonances, 
we use  Eq.~(\ref{eq:mesgaga}) to provide a first estimate of their contribution to the sum rule.  We 
show the results in Table~\ref{table_tensor}, both in the narrow width approximation and using a Breit-Wigner shape, assuming that the tensor mesons pre-dominantly contribute to $\sigma_2$, 
as is found by the experimental analyses of decay angular distributions~\cite{Mori:2007bu,Uehara:2008pf,etapi:2009cf,Abe:2003vn}. 

 By comparing Tables~\ref{table_ps} and \ref{table_tensor}, we notice that the contribution to the sum rule of the lowest isovector tensor meson composed of light quarks, $a_2(1320)$, compensates to around 70 \% the contribution of the $\pi^0$, which is entirely 
governed by the chiral anomaly. 
For the isoscalar states composed of light quarks,  the cancellation is even more remarkable, 
as the sum of  $f_2(1270)$ and $f_2^\prime(1525)$ cancels entirely, within the experimental accuracy, the combined contribution of the $\eta$ and $\eta^\prime$.   

\begin{table}[t]
{\centering \begin{tabular}{|c|c|c|c|}
\hline
& $m_M$  & $\Gamma_{\gamma \gamma} $   &  $\int ds\;  \Delta \sigma  / s$  \\
&  [MeV] &  [keV] &  [nb]  \\
\hline 
\hline
$\pi^0$ & 134.98   &  $(7.8 \pm 0.6) \times 10^{-3}$  &  $-195.0 \pm 15.0$   \\
\hline
\hline
$\eta$ &  547.85   &  $0.51 \pm 0.03$  &  $-190.7 \pm 11.2 $  \\
$\eta^\prime$ & 957.66   &  $4.30 \pm 0.15$  & $ -301.0 \pm 10.5$   \\
\hline
Sum $\eta, \eta^\prime$ & & & $-492 \pm 22$ \\
\hline
\end{tabular}\par}
\caption{Sum rule contribution of the lowest pseudoscalar mesons (last column). 
The experimental values of meson masses~$m_M$ and $2 \gamma$ decay 
widths $\Gamma_{\gamma \gamma}$ are  from Ref.~\cite{Amsler:2008zzb}.}
\label{table_ps}
\end{table}

\begin{table}[t]
{\centering \begin{tabular}{|c|c|c|c|c|}
\hline
& $m_M$  & $\Gamma_{\gamma \gamma} $   &  $\int ds\;  \Delta \sigma  / s$ & $\int ds\;  \Delta \sigma  / s$  \\
&  &    &  \rm{narrow res. }  & \rm{Breit-Wigner} \\
&  [MeV] &  [keV] &  [nb]  & [nb] \\
\hline 
\hline
$a_2 (1320)$ &  $1318.3$   &  $1.00 \pm 0.06$  &  $134 \pm 8$  &  $137 \pm 8$   \\
\hline
\hline
$f_2 (1270)$  & $1275.1 $   &  $3.03 \pm 0.35$  &  $448 \pm 52$  & $479 \pm 56$   \\
$f_2^\prime (1525)$  & $1525 $   &  $0.081 \pm 0.009$  &  $7 \pm 1$  & $7 \pm 1$   \\
\hline
Sum $f_2, f_2^\prime$ & & & $455 \pm 53$ & $486 \pm 57$ \\
\hline
\end{tabular}\par}
\caption{Sum rule contribution of the lowest tensor mesons. We show both results in the narrow resonance approximation (4th column) and using a Breit-Wigner parametrization (last column). 
The experimental values of meson masses $m_M$ and $2 \gamma$ decay 
widths $\Gamma_{\gamma \gamma}$ are  from Ref.~\cite{Amsler:2008zzb}.}
\label{table_tensor}
\end{table}

Besides the tensor mesons, the subdominant resonance contributions to the $\gamma \gamma$ total cross section arise from the scalar mesons 
$f_0 / \sigma(600)$, $f_0(980)$, and $a_0(980)$. A reliable estimate of the scalar mesons requires an amplitude analysis of the partial channels, see e.g.~\cite{Pennington:2008xd}. A future study will estimate more precisely the scalar meson helicity-zero contribution to the sum rule, and elaborate on the cancellation between the 
tensor mesons and the (pseudo)scalar meson contributions in the sum rule of 
\Eqref{s0rulesb}. Interestingly,  when going to the charm sector, 
the sum rule also implies a cancellation between the $\eta_c$ meson, whose contributions amounts to about  $-15.5$~nb,  and scalar and tensor $c \bar c$ states. 

We emphasize that our above analysis relies on a separation of the helicity-0 and helicity-2 cross sections using angular distributions of the decay products. To measure these cross-sections
and thus verify the sum rule directly, one needs polarized-beam colliders.

We conclude with a look at the new sum rules, expressed by \Eqref{s0rulesc}. These sum rules  completely determine the  constants $c_1$
and $c_2$, characterizing the low-energy photon
self-interaction, in terms of linearly-polarized fusion cross-sections. The sum rule for
$c_1+c_2$ can be viewed as the analog
of the Baldin sum rule for the sum of electric and 
magnetic polarizabilities~\cite{Baldin}.  The sum rule for  $c_1-c_2$
is unique to the $\ga\ga$ system.

To verify these sum rules in QED we recall that to leading order
all the ingredients are well known:
$c_1=\nicefrac{1}{90}\,  \al^2 m^{-4}$, $c_2=\nicefrac{7}{360}\, \al^2 m^{-4}$, while
the linearly polarized cross-sections can
 be found in the Appendix of \cite{Budnev:1974de}.
At the same time, these sum rules can be used to measure $c_1$ and $c_2$
through polarized $\ga\ga$ fusion experiments. 

In summary, we have presented a set of sum rules involving polarized 
total inclusive cross-sections of photon fusion and verified them in leading-order QED. 
One of them is the known superconvergence relation for the helicity-difference cross-section.
When applied to hadron channels, it reveals an intricate cancellation between the pseudoscalar- and tensor-meson contributions.
Two further sum rules determine the low-energy photon self-coupling in terms
of the integrals of linearly-polarized cross-sections for photon fusion.

\medskip
We thank Achim Denig and Miriam Fritsch for helpful discussions of the experimental situation.

 \end{document}